# A Cross-Dataset based Zero-Day Intrusion Detection System by Integrating Siamese Network and Reinforcement Learning

Md. Meheraj Hossain, Saumik Das Turja, Sibgatullah Tasnim, Md. Fahmid-Ul-Alam Juboraj, Muhammad Iqbal Hossain
Department of CSE, BRAC University, Dhaka, Bangladesh

***Abstract*—Zero-day threats are nascent for the Internet of Things (IoT) network security, which demands cognitive detection mechanisms that can identify emerging malicious behavior. Conventional intrusion detection mechanisms fail to generalize dynamic zero-day exploits within sophisticated IoT environments. This paper proposes a hybrid zero-day intrusion detection system using Siamese network-based anomaly correlation and reinforcement learning-based adaptive defense. Furthermore, the paper uses unsupervised machine learning classifiers over benchmark IoT datasets with the intention of detection of known attack types compared to unknown anomalies using distance-based similarity analysis to detect possible zero-day attacks. To facilitate adaptability, a Proximal Policy Optimization (PPO) reinforcement learning-based agent dynamically adjusts the defense policy with continuous feedback and optimization. Experimental evaluations demonstrate 99.28% training accuracy, 99.07% accuracy in unknown attack detection, and 93.94% zero-day detection ratio, confirming the convergence and stability of the model on datasets.The system offers a self-learning and extensible defense mechanism of IoT deployments by finding the right balance between precision, latency and false positives. This deep anomaly correlation with adaptive reinforcement learning is a firm base on which the next generation and autonomic cyber security solutions can take the reins as the zero-day threats keep changing their course.**

## I. INTRODUCTION

Zero-day attacks remain the most difficult security hazard, as they leverage flaws in software and hardware unknown to security vendors. This kind of vulnerability leaves developers almost no time to apply patches Satam et al., [7](2023). It is difficult for traditional signature based detection systems to manage continuously changing threats Pitre et al., [4] (2022),Perumal et al., [9] (2024). Attackers frequently employ sophisticated encryption, code mutation, and anti-analysis malware to bypass static defenses Al-Rushdan et al., [1] (2019), Abbas-Escribano and Debar [5](2023),Diloglu [2](2022). Although anomaly-based detection systems indicate viable solutions, it struggles with the high false-positive rate that deluges the security analysts with false alerts questioning the reliability of the system Kareem et al. [8](2024).

Reinforcement learning (RL) is able to scan and process big amounts of data by seeking out concealed trends and patterns; this is considered a shared limitation of static pattern detection systems [28]. The proposed model in this thesis combines unsupervised anomaly detection, cross-dataset Siamese analysis, and reinforcement learning based adaptive response that can recognize previously unknown attack patterns in an IoT network. This model is aimed at lowering false positive rates while exploring the implications of Reinforcement learning (RL) in threat detection. To design the most efficient architecture, this research surveys various unsupervised models, different distance measures for Siamese analysis as well as multiple reinforcement-learning algorithms. The Rl module uses a customized reward function designed to optimize performance/adaptability trade-off. Lastly, impact of four hyperparameters on overall system performance was documented while tuning.

## II. LITERATURE SURVEY

Wahed et al. [12] (2025) proposed an AI-enhanced IDS employing RNN, LSTM, and GRU models for sequential network traffic analysis. The architecture processes real-time data with approximately 2-3 million records per attack type, achieving 98.5% accuracy and maintaining false positive rates below 8%. The system integrates Random Forest optimization and NLP techniques for security report analysis, enabling automatic threat response through session disconnection and security alerts. However, computational demands limit deployment in resource-constrained environments, and continuous retraining is essential for evolving threats.

Zaki et al. [11] (2024) developed an ensemble IDS for IoT zero-day attack detection using sparse random projection with KNN, XGBoost, and SGD classifiers. Evaluated on CIC-DDoS2019 and CICIDS2017 datasets, the system achieved 99.91% accuracy with 0.22-second detection time through GridSearchCV hyperparameter optimization. While effective for IoT environments, the approach's generalizability to broader network infrastructures remains limited due to dataset-specific training. Sarhan et al. [6] (2023) introduced a zero-shot learning framework addressing signature-based NIDS limitations through semantic attribute learning. The methodology trains ML models on known attacks (phishing, DDoS, XSS, SQL injection) to generalize zero-day attack detection. Using MLP and Random Forest models, the system achieved 95% detection rate for generic attacks but struggled with complex reconnaissance and exploit patterns (90% Z-DR). The framework requires sufficient training data and may face accuracy degradation in dynamic network environments. Diloglu [2] (2022) implemented supervised and unsupervised deep learning models for zero-day detection across KDD'99, NSL-KDD, UNSW-NB15, and CIC-IDS2017 datasets. The approach utilized artificial neural networks and autoencoders with batch normalization, achieving 97% accuracy on CIC-IDS2017. While supervised models demonstrated superior performance, unsupervised approaches showed limitations on legacy datasets, highlighting challenges with high-dimensional data and dataset relevance. Kareem et al. [8] (2024) utilized Neural Transformers for zero-day detection by processing network packet sequences (packet size, protocol type, addresses). Testing on CICIDS 2017 dataset, the transformer model achieved 96% accuracy and 94% precision with 25-millisecond latency, outperforming traditional ML approaches. Despite superior performance, computational complexity limits its resource-constrained deployment, requiring model quantization and knowledge distillation optimization.

Nkongolo et al. [3] (2022) developed a cloud-based zero-day detection methodology using Genetic Algorithm optimization with SVM, Naive Bayes, and Random Forest classifiers. Deployed on AWS S3 and SageMaker platforms, the Random Forest model achieved 99.6% accuracy on CAIDA and UNSWNB-15 datasets. The Genetic Algorithm enhanced weak classifier performance through ensemble learning and feature selection, though individual classifier instability and limited training samples remained constraints. Roopak et al. proposed an unsupervised IDS for IoT zero-day DDoS detection using Gaussian Random Projection for dimensionality reduction (87 to 25 dimensions). The ensemble method combines K-means, GMM, and SGD-OCSVM through hard voting, achieving 94.5% accuracy and 94.3% F1-score on CIC-DDoS2019. While effective for high-dimensional unlabeled data, the ensemble architecture increases computational complexity and requires extensive hyperparameter tuning.

Abbas-Escribano and Debar [5](2023) implemented an enhanced high-interaction honeypot system based on MITRE ATTACK taxonomy for attack analysis. The virtualized multi-level architecture recorded 1.5 million events over 17 days, successfully logging attacker behavior and generating IoCs while preventing deeper system penetration. However, the relatively short exposure period and limited advanced attacker attraction capabilities constrained comprehensive threat analysis.

## III. Methodology

The overall objective of this work is to design an intelligent intrusion detection system capable of identifying both known and unknown threats. To achieve this, the design process is organized into multiple consecutive stages to successfully screen out unseen anomalies.The method uses unlabeled traffic data of two large IoT datasets therefore, it is able to identify abnormal patterns without supervised classification or manual labeling. The framework aims at providing dynamic security against cyber threats. Firstly, unsupervised anomaly detection models are trained on IoT attack datasets with only semantically matched shared features retained. Anomalies extracted from both unsupervised anomaly detection models are then compared and based on AUC score, spread and F1 score of three distance measures—Euclidean, Manhattan, and Cosine distances; best threshold possible is selected for identifying dissimilar and rare attack patterns. This measure is selected for ultimate anomaly extraction based on this comparison for optimum detection fidelity. Filtered anomalies serve as training data for a reinforcement learning agent based on an Actor-Critic architecture and a value based method. The RL agent is trained only on attack data with the goal to learn adaptive defense actions such as selective blocking and alerting. The PPO learning model is tested with controlled injection of two different data sets: (1) the training set, to ensure the agent's attack-response behavior is retained after learning, and (2) an entirely unknown set, to test its unseen attack detection capacity. To conclude, by integrating unsupervised anomaly detection, strict cross-dataset anomaly validation, and reinforcement learning–driven response mechanisms, the pipeline moves proactive defense forward for zero-day attack detection and mitigation in IoT networks.

**Dataset** For this paper two IOT datasets, CIC-BCCC NRC TabularIoT 2024 and CIC IoT 2023; both from the official Canadian Institute for Cybersecurity (CIC) repository, publicly accessible to everyone for academic and research purposes were used.

**CIC-BCCC-NRC-TabularIoTAttacks-2024 dataset:** This is an augmented labeled and tabular collection of nine common IoT datasets containing over 1 million benign and malicious network traffic data and includes detailed feature extraction using CICFlowMeter.

**CIC IOT 23 dataset:** A collection of 11 different types of DDOS attack, 1 brute force attack, 2 kinds of spoofing attack, 4 kinds of DOS attack, 5 kinds of Recon attack, 6 kinds of Web-based attack and 3 kinds of Mirai attack are captured in individual CSV files with detailed network attributes including flow statistics, packet details, TCP flags and protocol behaviors. This dataset was selected due to its diversity of realistic attacks and well balanced benign and malicious data.

### Data Cleaning

The data cleaning procedure ensured dataset consistency and suitability for unsupervised anomaly detection. Infinite values were replaced with NaN and subsequently removed alongside rows with missing values to maintain dataset validity. Duplicate entries were eliminated, and z-score analysis was used to focus on significant values. The training dataset was balanced, requiring no further balancing techniques for this work. For reinforcement learning, transitions containing infinite values or outliers were filtered, and normalization was applied to state features, rewards, and logged metrics to ensure stable training dynamics.

Feature scaling using StandardScaler normalized attributes to zero mean and unit variance, preventing features with large magnitudes from dominating. Manual feature selection retained only meaningful features, reducing dimensionality and noise, thereby enhancing anomaly detection efficiency. Prior to model input, irrelevant columns such as IP addresses and timestamps were removed, with categorical labels transformed to numeric form for compatibility.

Data integration merged two IoT-centric datasets—CIC IoT Dataset 2023 and CICBCCCNRCTabularIoTAttacks2024—after aligning overlapping features and harmonizing attribute encodings (e.g., Protocol Type). Consistent preprocessing steps ensured comparability for cross-dataset anomaly detection. Dataset labels were retained for evaluation but excluded during unsupervised training.

Dimensionality reduction and distance-based anomaly filtering, using metrics like Euclidean distance, further refined feature sets and isolated unusual anomalies of interest in zero-day detection, balancing computational efficiency with analytic focus.

### Unsupervised Anomaly Detection

This study compared performance of five unsupervised machine learning algorithms—PCA-Based Anomaly Detection, Standard Autoencoder, Isolation Forest, One-Class SVM, K-means Clustering in identifying anomalies in IoT network traffic. Two IoT datasets—CIC IoT Dataset 2023 and CIC BCCC NRC Tabular IoT Attacks 2024 were used in this phase.

The CIC-IOT2023 dataset was divided into a training set containing 1,000,000 benign samples and testing with about 4,000,000 samples containing both benign and attack traffic.The Attack type distribution of CIC-IOT2023 is shown in figure 1.

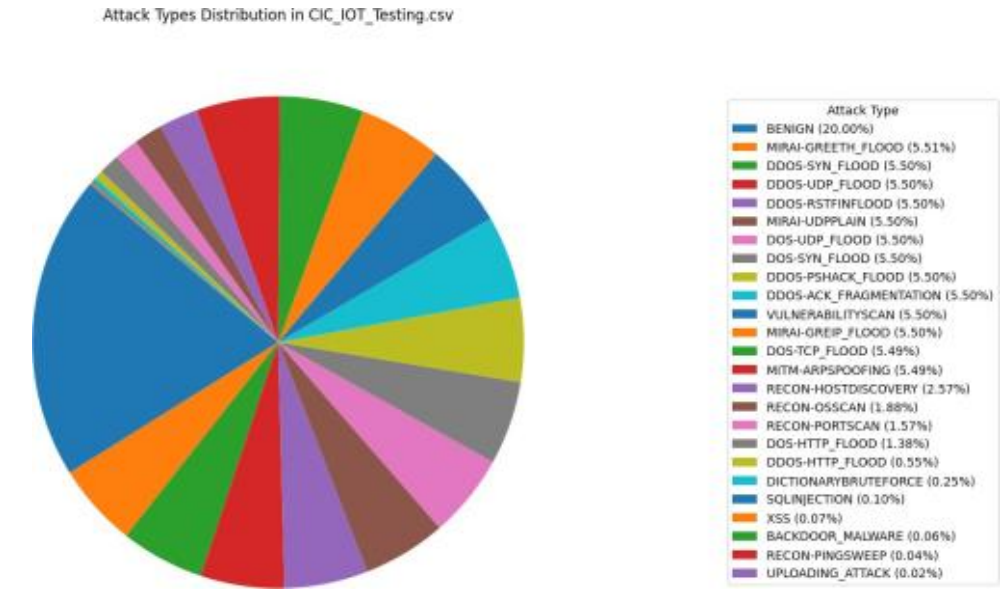

Fig. 1. Attack types distribution in CIC-IOT23

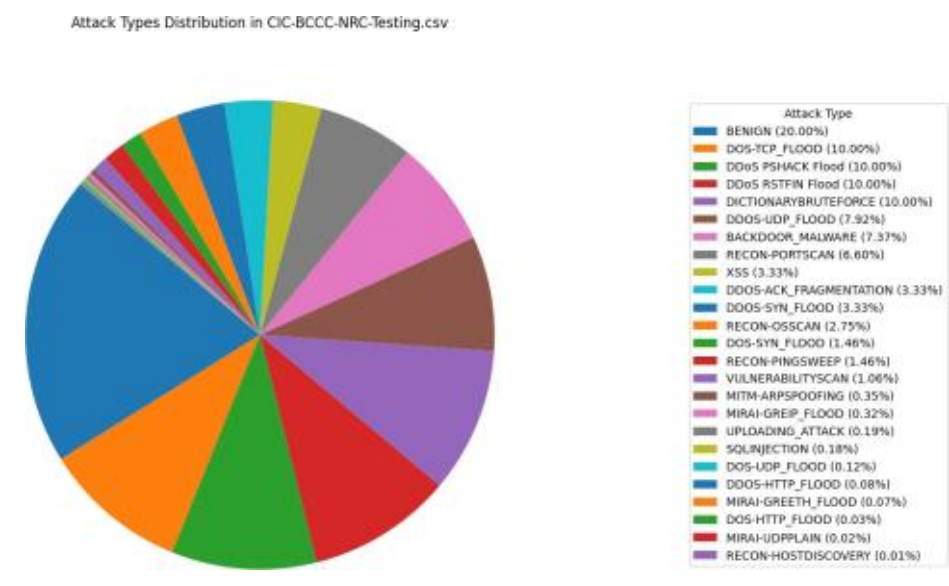

Fig. 2. Attack types distribution in CIC-BCCC-NRC-TabularIoTAttacks-2024

The CIC-BCCC-NRC-Tabular IoT Attacks-2024 is more detailed, sparse in its feature space of more than 80+ traffic features. Such features are detailed flow metrics, packet header statistics, timing information, and fine-grained flag counts, as obtained with advanced network flow analyzers such as CICFlowMeter. The Attack types distribution in CIC-BCCC-NRC-TabularIoTAttacks-2024 is shown in figure 2.

Elaboration included renaming columns to be consistent, combining similar features and ensuring categorical variables, like the protocol types, were aligned across the datasets. This enabled us to develop a consistent feature space for both datasets, necessary to support cross-dataset comparison. After thorough preprocessing and harmonization the semantically similar features recognized between the two datasets were approximately 20—making a subset of common features later used for comparison in cross-dataset and modeling transferability. Using these, we derived parity testing sets, to conduct performance and generalizability assessment of selected models consistently on the two datasets. Attack classes are withheld from both datasets to be used exclusively for testing to evaluate generalization and unseen attack detection capacity.

**CIC IOT Dataset2023:**

For this particular dataset Standard autoencoder was selected after evaluating performance of PCA-Based Anomaly Detection, Isolation Forest, One-Class SVM and K-mean and finding Standard autoencoder to be the best fit. Standard autoencodes have the ability to reconstruct normal data patterns and significant deviations from learned patterns are understood to be anomalous data.

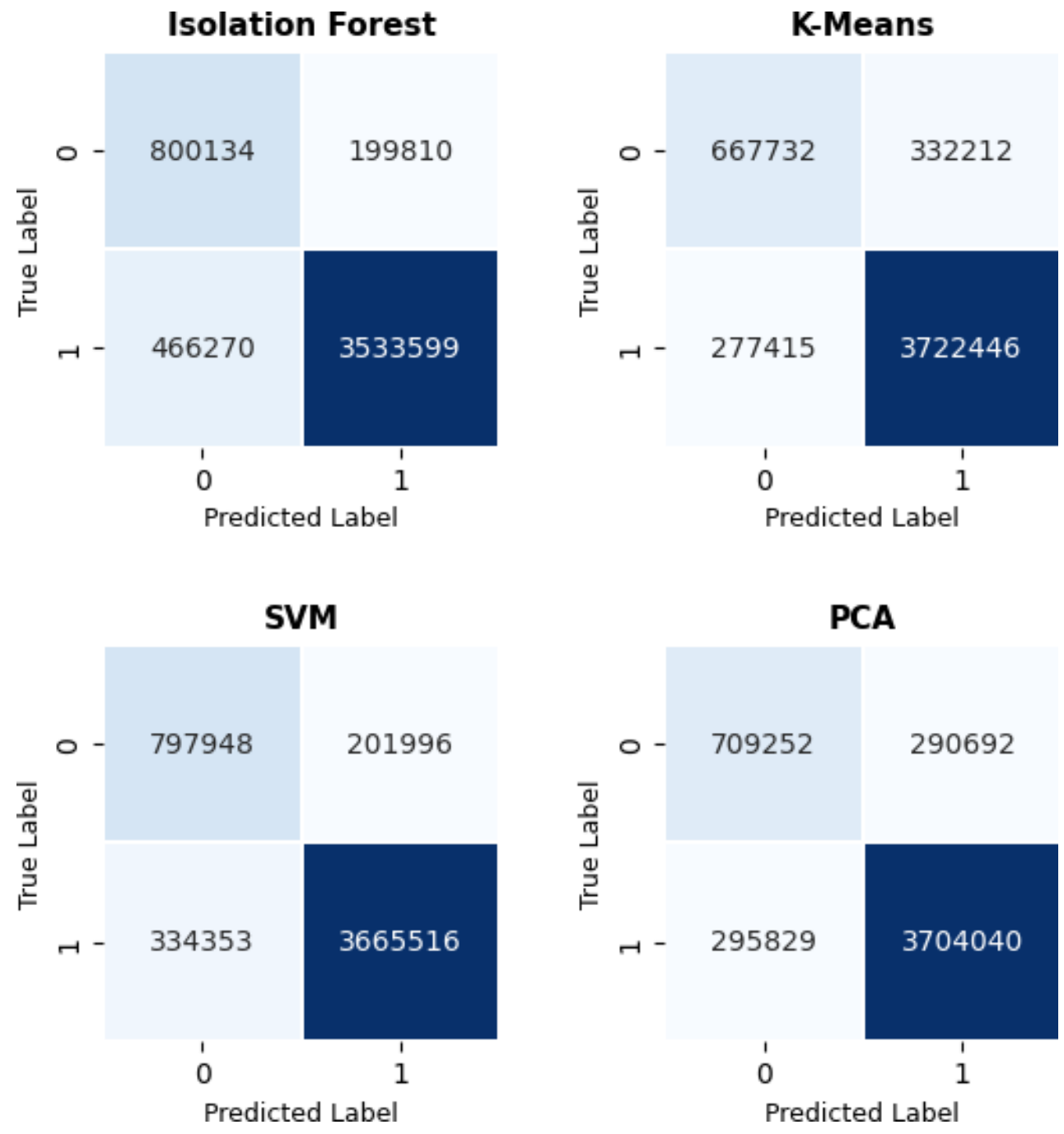

Fig. 3. Confusion matrices of Isolation forest, SVM, Kmean and PCA of CIC IOT 2023

Figure 3. are four confusion matrices for four unselected algorithms showing the four Confusion Matrix Components. These four outcomes collectively describe the performance of a classification model. In the case of Isolation Forest—800,134 (TN), 199,810 (FP), 466,270 (FN), 3,533,599 (TP), are the four Confusion Matrix Components with an accuracy of 88%.

TABLE I
EVALUATION METRICS OF ALL THE FIVE UNSUPERVISED MODELS OF CIC-IOT-2023

| Model | Accuracy | F1-Score | Recall |
|---|---|---|---|
| PCA | 0.88 | 0.87 | 0.88 |
| Autoencoder | **0.89** | **0.88** | **0.89** |
| Isolation Forest | 0.88 | 0.88 | 0.88 |
| One-Class SVM | 0.89 | 0.88 | 0.89 |
| K-means | 0.88 | 0.88 | 0.88 |

For CICIOTDataset2023, the priority was the overall F1/accuracy and attack detection power, in particular if we look at Table I. The Standard Autoencoder algorithm outperforms other algorithms with highest accuracy (0.89, tied with One-Class SVM, slightly better than others), F1-score and recall.

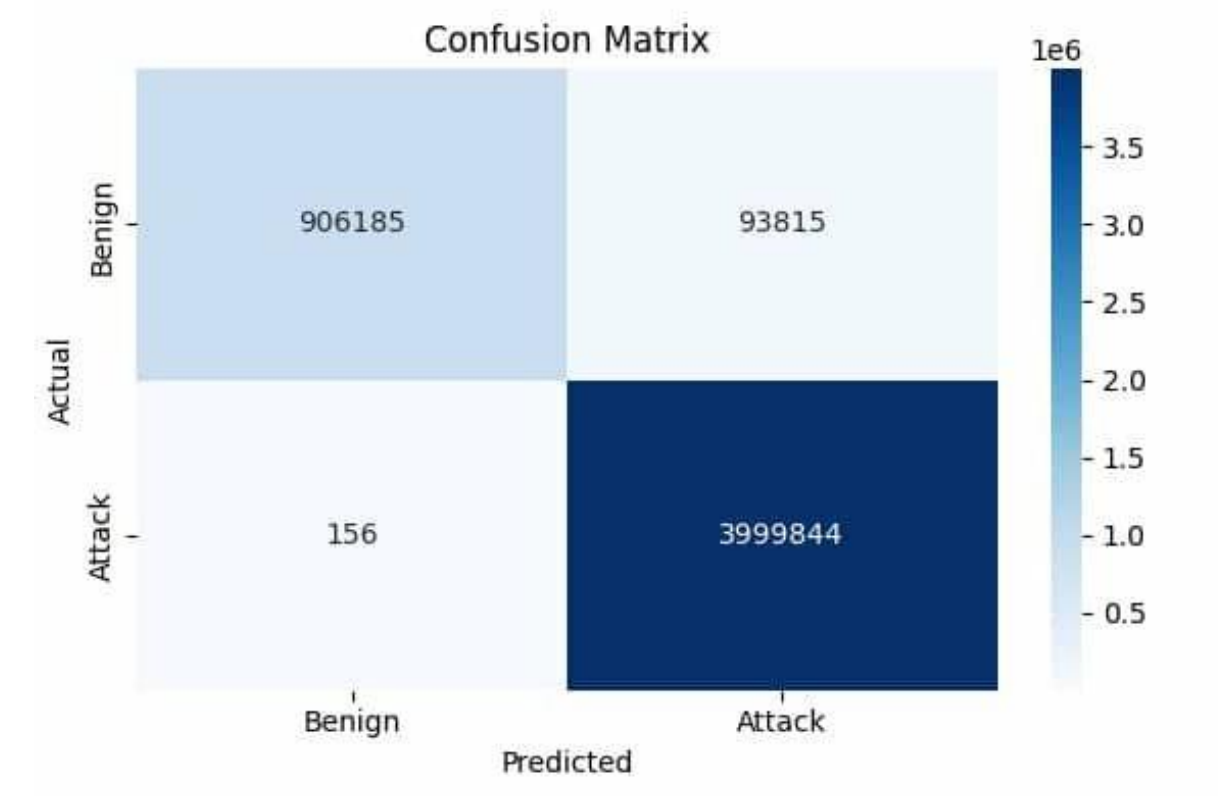

Fig. 4. Confusion matrix of Selected Model (Autoencoder)

In figure 4, the confusion matrix of standard autoencoder is shown.Here,true Negatives (650315 correct Benign),False Positives (349629 misclassified Benign),False Negatives (210609 missed Attacks) and True Positives (3789252 correct Attacks). Total: 4999805 samples. Accuracy=89

**CIC-BCC-NRC-TabularIotAttacks-2024:**

Principal Component Analysis (PCA) for the CICBCCC-NRCTabularIoTAttacks2024 was selected after evaluating performance of Isolation Forest, One-Class SVM, K-mean and Standard Autoencoder. It projects a high- dimensional dataset onto a set of novel, independent features known as principal components, retaining useful information This is particularly useful for anomaly detection.

In figure5, are four confusion matrices for four unselected algorithms showing the four Confusion Matrix Components. These four outcomes collectively describe the performance of a classification model. In the case of Isolation Forest— 796518 (TN), 203482 (FP), 7126 (FN), 3992874 (TP), are the four Confusion Matrix Components with an accuracy of 88%.

For CIC BCCC NRC TabularIoTAttacks-2024, Autoencoder, K-Means, and PCA all produce nearly identical results as shown in Table II.

**Performance Evaluation of the selected model**:

Figure 6 is the confusion matrix to selected model (PCA) where True Negatives (907493 correct Benign), False Positives

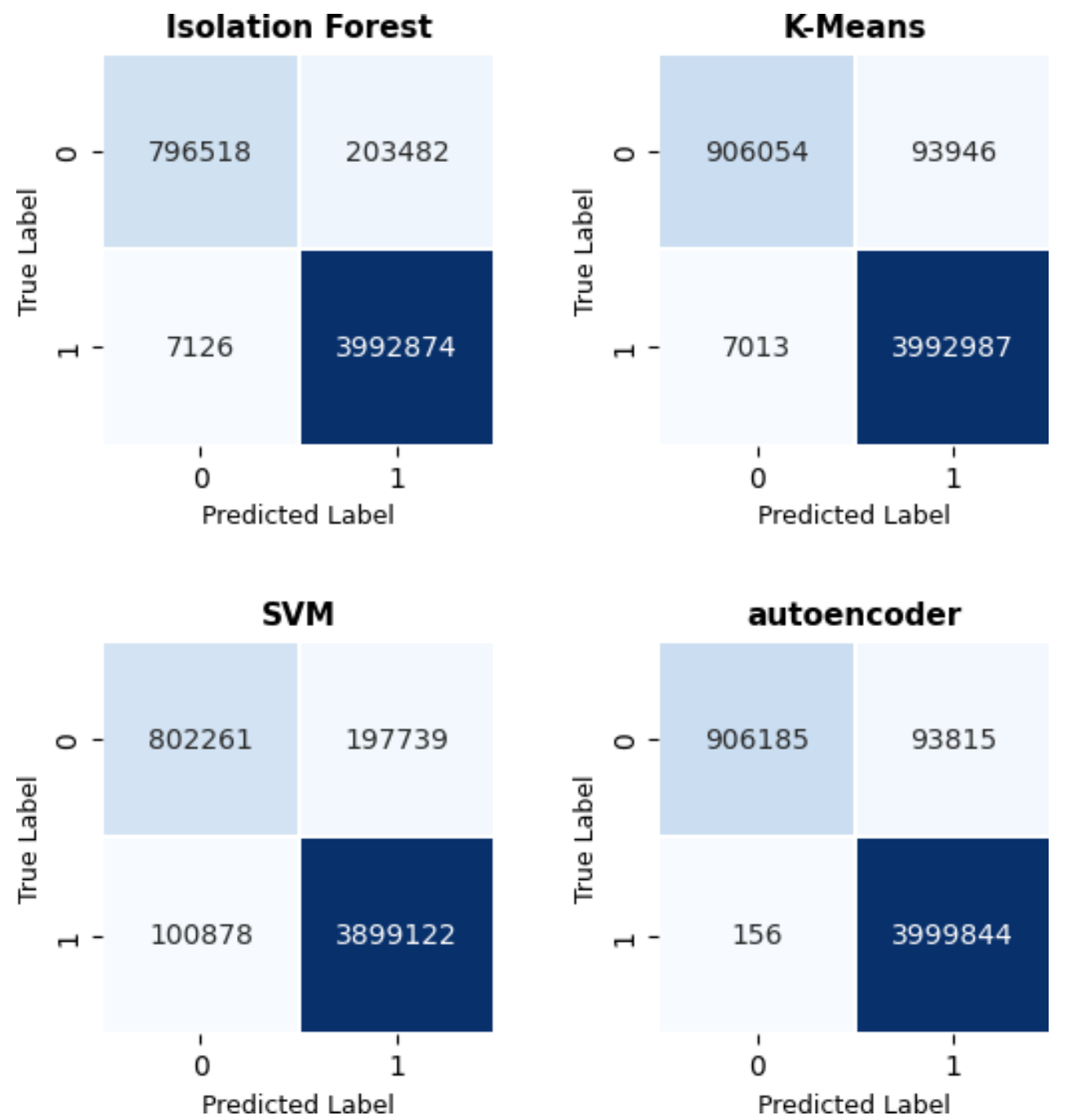

Fig. 5. Confusion matrices of Isolation forest, SVM, Kmean and autoencoder of CIC-BCC-NRC-TabularIotAttacks-2024

TABLE II
EVALUATION METRICS OF THE FIVE UNSUPERVISED ANOMALY DETECTION ALGORITHMS

| Model | Accuracy | F1-Score | Recall |
|---|---|---|---|
| PCA | **0.98** | **0.98** | **0.98** |
| Autoencoder | 0.97 | 0.98 | 0.98 |
| Isolation Forest | 0.95 | 0.95 | 0.95 |
| One-Class SVM | 0.94 | 0.96 | 0.97 |
| K-means | 0.97 | 0.98 | 0.98 |

(92507 misclassified Benign), False Negatives (5319 missed Attacks) and True Positives (3994681 correct Attacks). Total: 500,000 samples. Accuracy=98%

**Cross-Dataset Anomaly Filtering via Siamese Architecture:** High-fidelity zero-day anomaly isolation using a Siamese Network is the next critical phase after respective dataset anomaly extraction from PCA and standard autoencoder modules. The Siamese network serves as a cross-dataset comparator as well as a principal filter to isolate the anomalies most likely to represent truly novel attack behaviors.The primary objective of this phase was to distill these outputs into a high-fidelity dataset containing only the most dissimilar and rare anomalies.

Each anomaly pair was scored with all three metrics, and the distance functions were evaluated for their effectiveness in

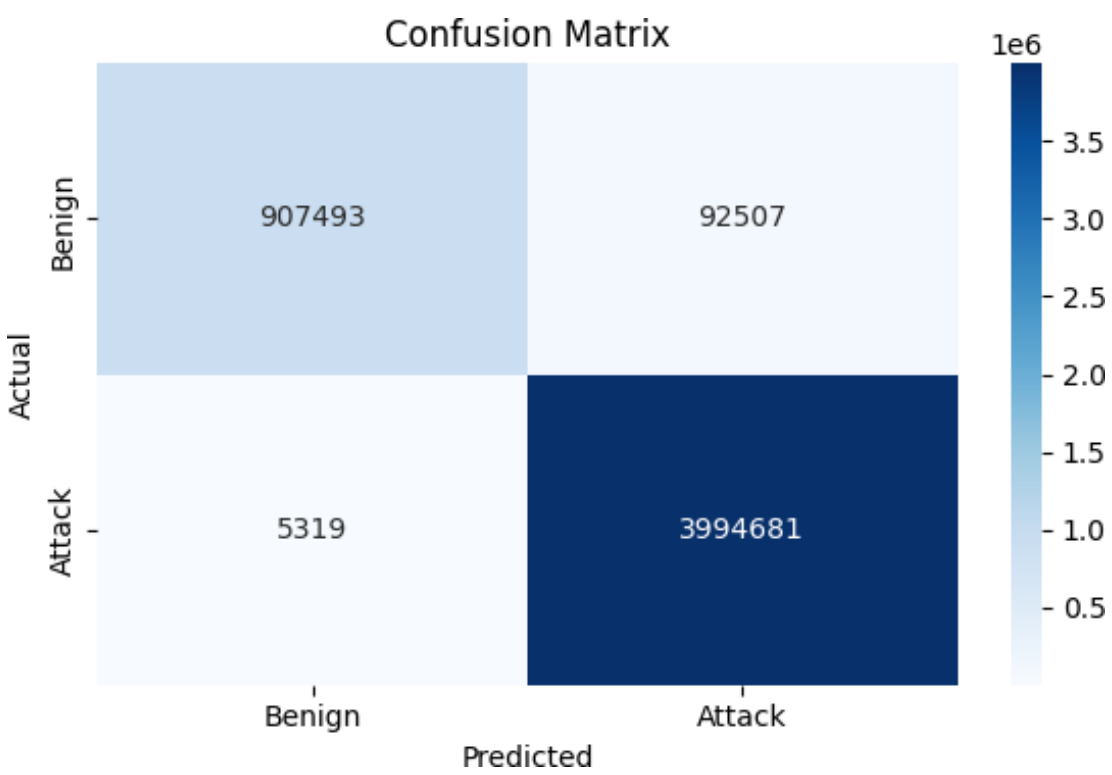

Fig. 6. Confusion Matrix of selected model (PCA)

identifying rare, high-risk anomalies. The evaluation metrics of the distance metrics models are shown in table III

TABLE III
EVALUATION METRICS OF THE THREE DISTANCE-BASED SIMILARITY MODELS

| Metric | AUC | F1-Score | Spread |
|---|---|---|---|
| Euclidean | **0.5048** | **0.3323** | **170026.6319** |
| Manhattan | 0.5046 | 0.3323 | 170022.4141 |
| Cosine | 0.4985 | 0.3320 | 0.0778 |

Every distance metric underwent a full ROC evaluation. Euclidean distance yielded the highest AUC (0.5048), F1 score (0.3323), and greatest “spread” (170,026), meaning it best maximized the separation between shared and truly unique attacks.

In figure 7,the AUC score is approximately 0.5 since the dataset was very imbalanced.Attack type count was significantly higher than benign count.

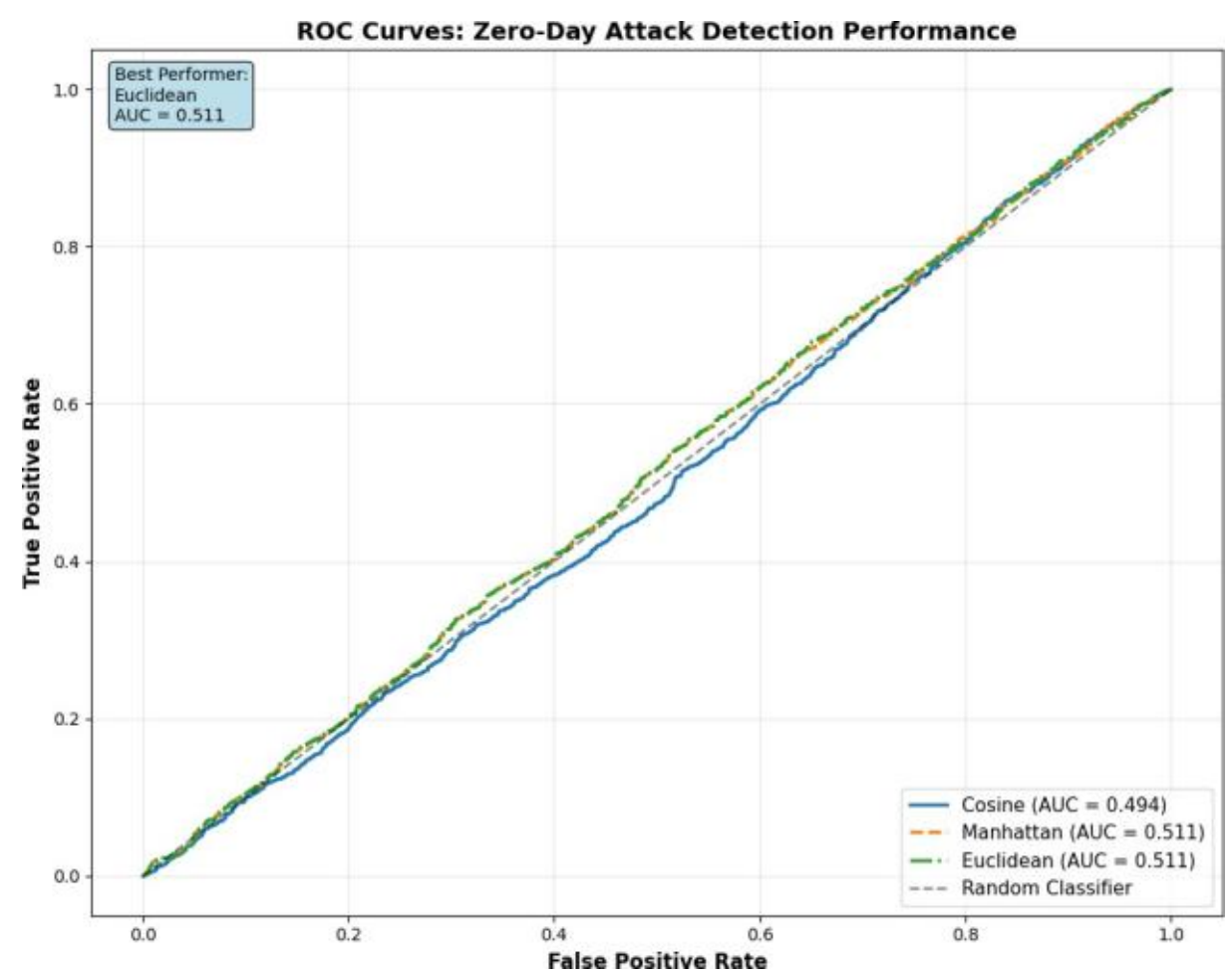

Fig. 7. ROC curve of the three distance metrics models

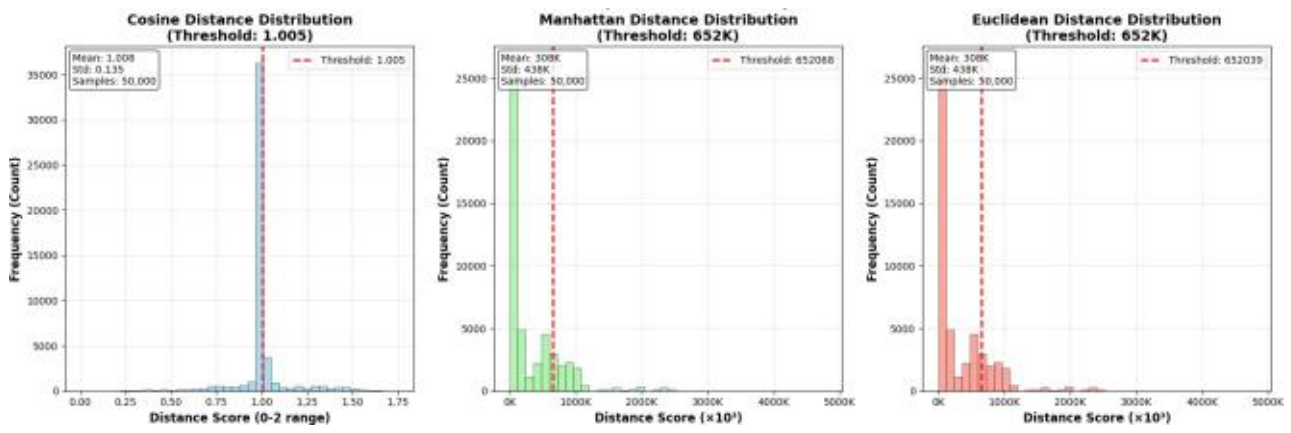

Fig. 8. Distance Metrics Comparison

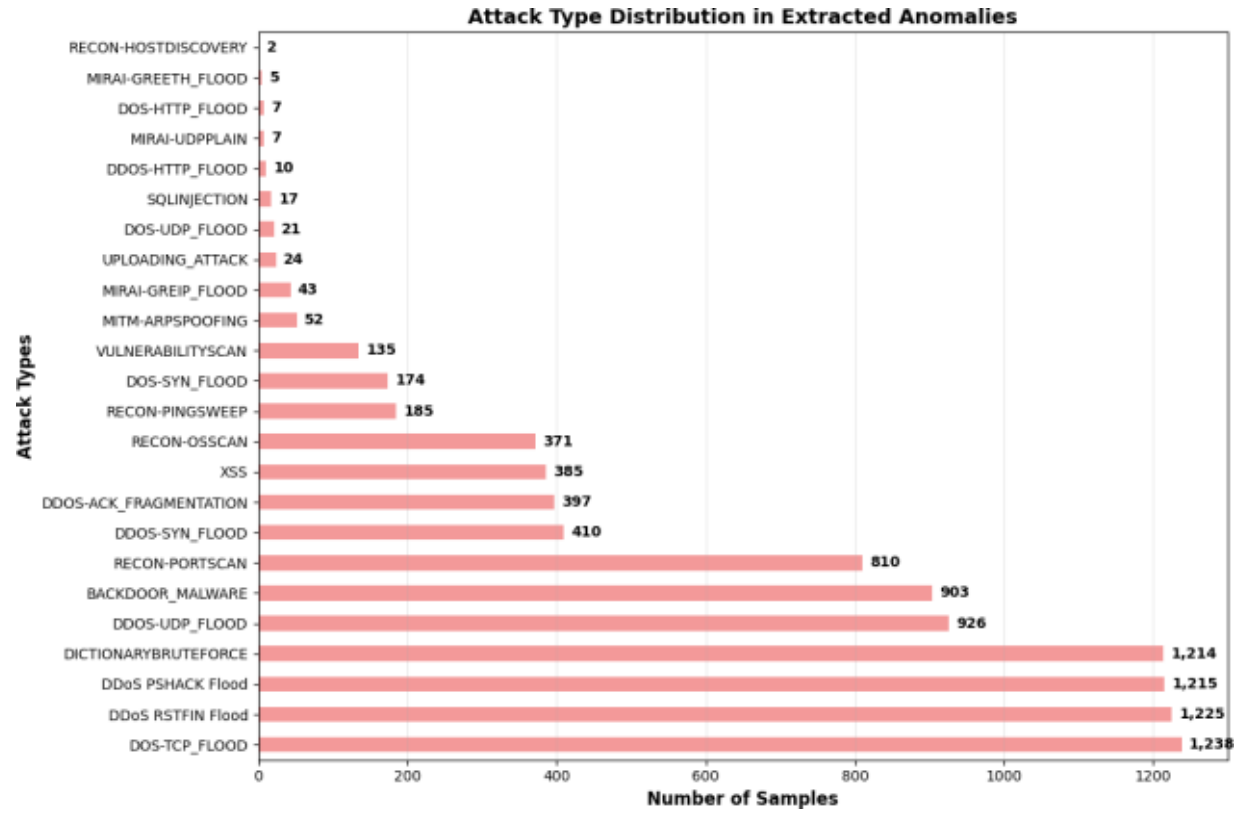

Fig. 9. 24 types of attacks in extracted anomalies

In figure 8 To detect the potential novel attacks, this three-panel histogram compares the Cosine, Manhattan, and Euclidean distances between scaled 2023-2024 network anomaly features (50K samples).Abrupt spike at 1.0 reflecting high direc- tion similarity with very few departures above threshold 1.004 is evident through the cosine (mean 1.006, std 0.131). Between 0-350K, Manhattan (mean 310K, standard deviation 442K) and Euclidean (identical statistics) possess right-skewed decays with tails indicating possible anomalies at threshold values of 653K.Others emphasize magnitude spreads, while cosine excels in compactness and extracts about 10,000 rare anomalies for varied RL training (e.g., 57% DOS-TCP FLOOD).

**Final Anomaly Extraction**: Applying the optimal metric and threshold produced the finalized anomaly set for downstream RL training. The deliberately skewed distribution confirms that the pipeline successfully isolated rare, novel, and high- consequence threats. These labeled outputs, by attack type and sample count, become the foundational inputs for reinforcement learning–driven defense policy optimization.24 types of attacks in extracted anomalies is shown in figure 9

**Reinforcement Learning**

The unique anomaly dataset produced is used to train PPO, Deep Q-Network (DQN), and SAC models in order to develop a defense agent that learns adaptive strategies, such as selective blocking or alerting, without requiring labeled data. The reinforcement learning will be to support adaptive responding in real time so that the system will learn and adapt automatically to new patterns of threat with no human intervention/or signature updates.The linked architecture ensures the model identifies suspected risks while also effectively labels abnormal behaviour— that could potentially represent new or develop- ing attack patterns. These modifications introduce flexibility, extending intrusion detection capacities to changing network environments In short, the approach is crafted for detection, learning and adaptability.

## IV. RESULTS AND DISCUSSION

### A. *Performance Evaluation*

Performance evaluation for the proposed model will be discussed in this section, based on the computational results of the best-performing algorithmic variant—PPO. The results demonstrate model adaptability. A simple reward system is used to direct the agent rewards with +1 for every correct detection (both seen and unseen) while an incorrect detection is penalized by -1.

Regular performance metrics are used for evaluation, in addition to the zero-day detection rate — i.e., the proportion of newly discovered attacks correctly identified, which assesses generalization ability. The actual threats are detected in terms of sensitivity (recall), and the ROC curve is used to measure classification quality by identifying the area under the receiver operating characteristic curve (AUC). In table IV, the combined evaluation metrics and RL hyperparameters for the proposed model are shown.

TABLE IV
COMBINED EVALUATION METRICS AND RL HYPERPARAMETERS FOR THE PROPOSED MODEL

| Metric | Value |
|---|---|
| Training Accuracy | 99.28% |
| Unseen Accuracy | 99.07% |
| Zero-Day Detection | 93.94% |
| Latency | 0.50 ms/sample |
| Learning Rate | $1 \times 10^{-3}$ |
| Gamma | 0.99 |
| Clip Epsilon | 0.15 |
| Episodes | 300 |

### B. ***Testing Methodology***

Testing methodology adopts a strict dual-dataset approach using the training csv, consisting labeled attacks and the unseen normalized csv, consisting of unseen attacks. An 80th percentile distance-based classification, determined to be optimum via empirical investigation, is used for threshold tuning. Distance-based ground truth labeling, for which points above or at the threshold distance are labeled zero-day, is used for evaluation. Performance is evaluated on both seen and unseen csv simultaneously, using an aggregated-scoring method that gives training performance 40% weight and unseen performance 60% weight. For repeatability, several independent runs are carried out for validation purposes.

### *C. Classification Excellence*

PPO configuration achieved stellar classification performance across all datasets with a training accuracy of 99.28%, novel attack detection accuracy of 99.07% demonstrating strong generalizability and an overall performance of 99.16%. The system proved good generalization with a training shift of only 0.21%. Furthermore, a training precision of 99.30% with an F1-score of 99.29%, validating accurate recognition and minor false classifications. Unseen precision was high at 99.16% with an F1-score of 99.10%. The false positive rate on previously unseen samples was a modest 2.21%, which reflects a small number of false alarm events.

### *D. Operational Efficiency*

The PPO model latency stayed at 0.50 milliseconds/sample, safely within real-time operational limitations. Memory consumption was also 203.56 MB—well within the operable limit. The model converged to 90 percent accuracy within 42 episodes.

### *E. Zero-Day Detection*

The PPO arrangement also scored remarkably zero-day detection accuracy with a 93.94 success rate. In addition, the training and unseen ROC AUC was 99.83% and 99.69% respectively, which confirms the model's high classification capacity and stability. Figure 10,11,12 shows the combined score of the Top 10 configurations, Optimal Configuration of the Learning Curve and summary of optimal configurations performance.

### *F. Analysis of Design SOlutions*

Ablation study was conducted to fine-tune four hyperparameters that allow optimal performance for PPO configuration.

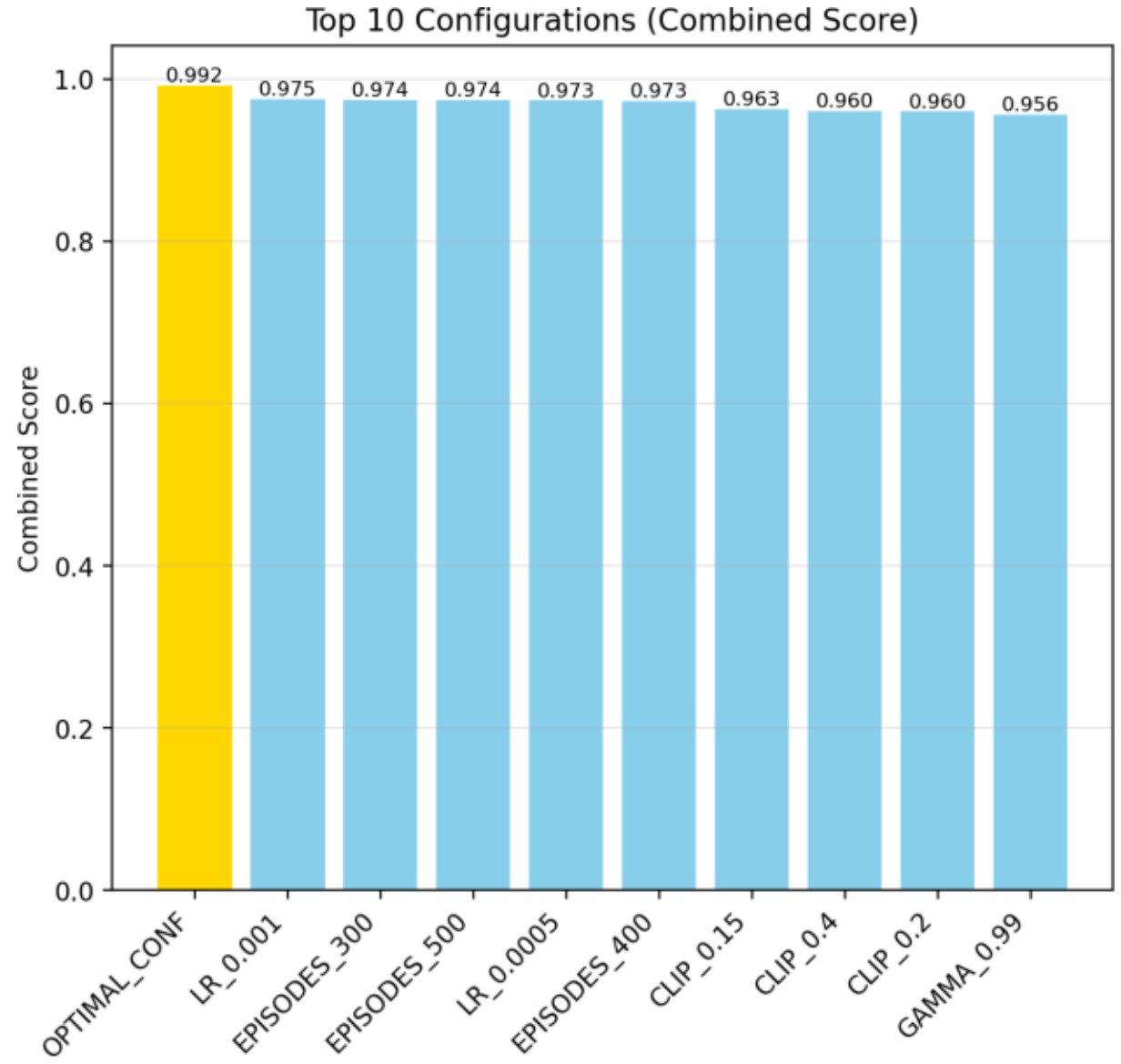

Fig. 10. Combined Score of the Top 10 Configuration

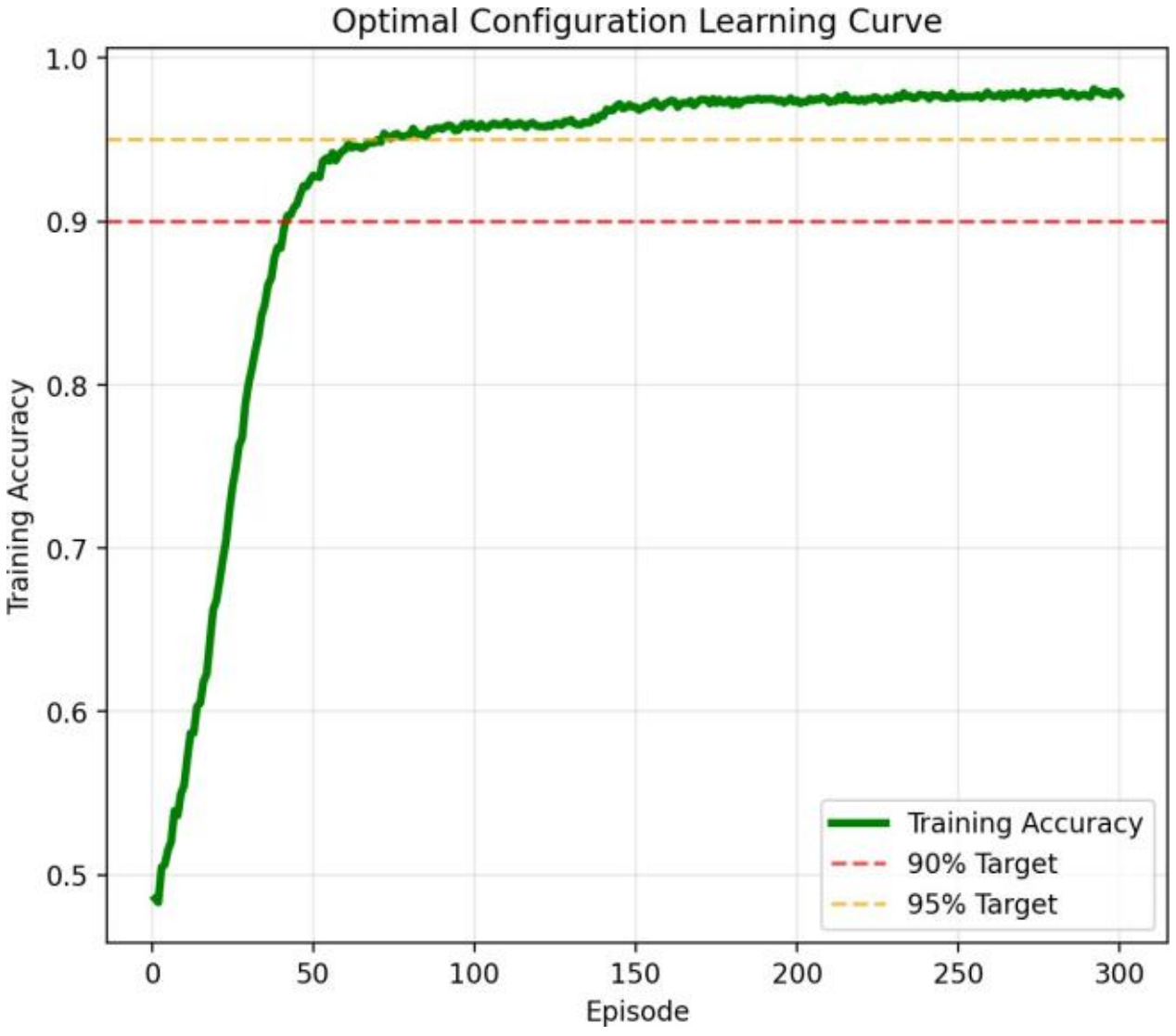

Fig. 11. Optimal Configuration of the Learning Curve

Ablation Study Methodology

The parameters of interest are the learning rate($\alpha$), which regulates the step size when updating policies; gamma ($\gamma$), which sets the discounting rate of future rewards; clip epsilon ($\varepsilon$), which provides stability for PPO policy updates when clipping; and training episodes, which identifies iterations required for the model convergence.

In this experiment a sequential optimizing method—which optimizes only one parameter at a time is used and the optimum values are carried across to the next iteration.A combined score was used to evaluate performance, weighted as 40% training accuracy and 60% unseen accuracy to emphasize generalization. Uniformity was ensured by maintaining the same controlled environment on all configurations. Statistical confirmation was achieved by running the results repeatedly to ensure reproducibility.

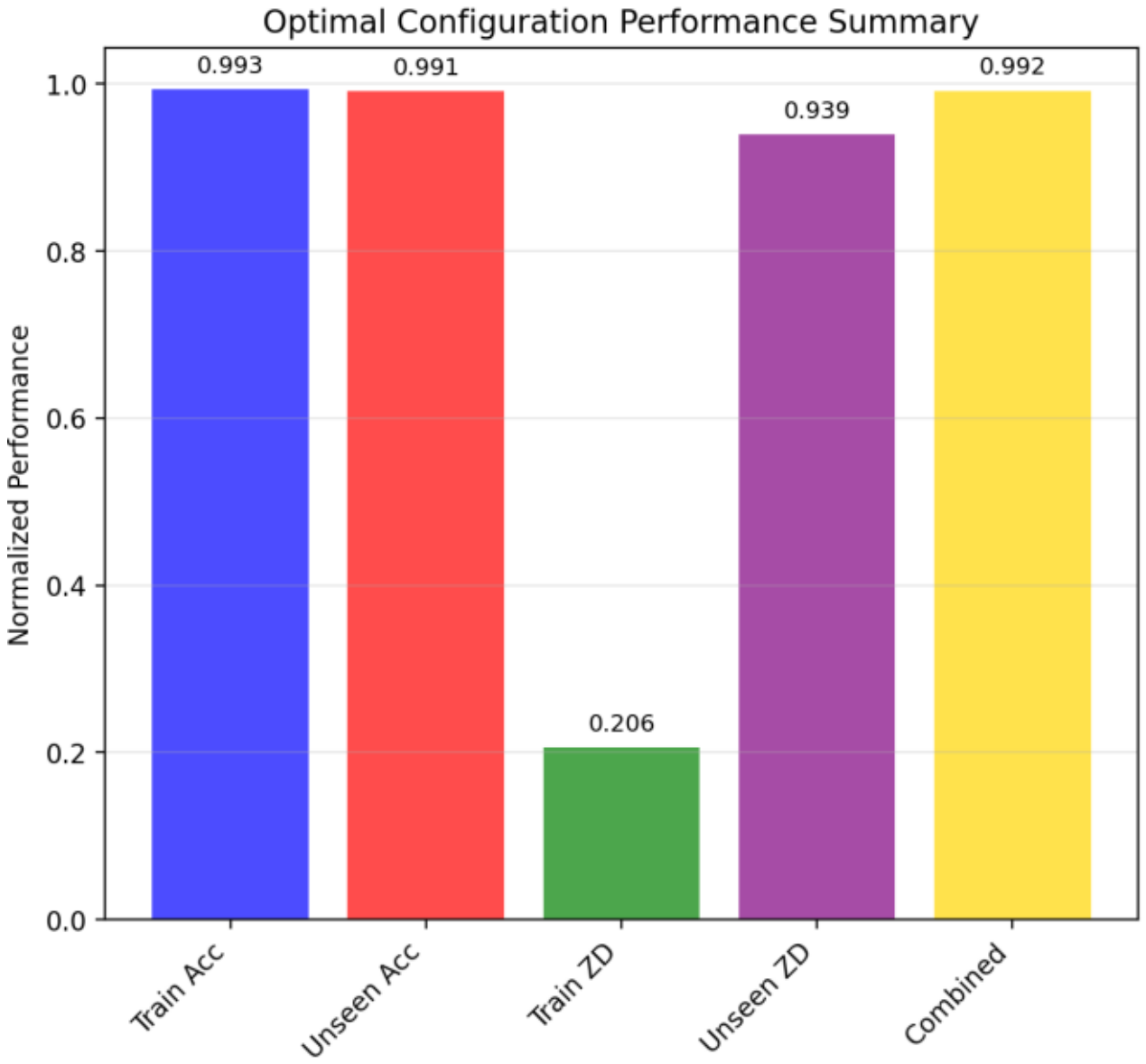

Fig. 12. Summary of the Optimal Configuration Performance

## G. *Learning Rate Optimization Analysis*

Figure 13 shows the analysis of the Learning Rate Optimization which provides accessible information on the stability and convergence nature of policy gradients. The range for evaluation was

$$[5 \times 10^{-5},\ 1 \times 10^{-4},\ 3 \times 10^{-4},\ 1 \times 10^{-3},\ 5 \times 10^{-3}]$$

for which the optimum was experimentally found to be $1 \times 10^{-3}$, giving a combined score of 0.975, corresponding to a performance impact of 0.233—the greatest of all parameters.

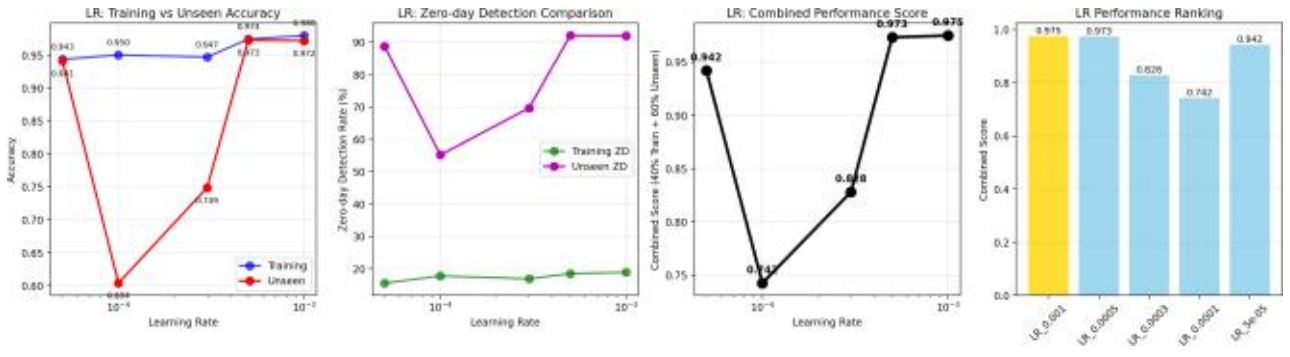

Fig. 13. Learning Rate Optimization Analysis

## H. *Gamma (Discount Factor) Optimization*

Figure 14 shows optimization of the Discount Factor which highlights short/long term reward optimization trade-offs. Test range values were [0.9, 0.95, 0.99, 0.995, 0.999], for which the optimum value 0.99 resulted in a cumulative score of 0.956 and performance impact of 0.067—the least of all parameters.

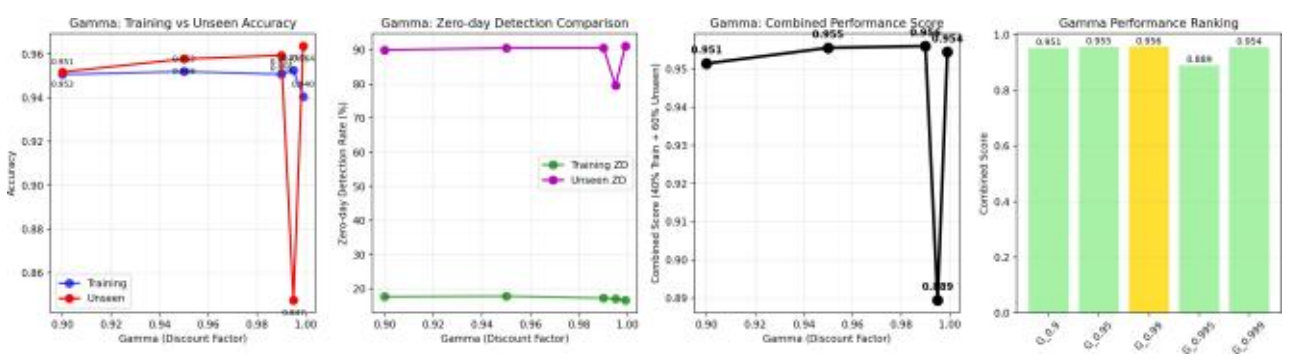

Fig. 14. Gamma (Discount Factor) Optimization

## I. *Clip Epsilon Optimization*

The range of the clipping epsilon range [0.1, 0.15, 0.2, 0.25, 0.3] was measured where 0.15 was experimentally found optimum with combined score of 0.960 and performance impact of 0.130. Conservative clipping (0.1 or less) was restricting policy changes—resulted in slow learning, whereas aggression clipping (0.25 or above) caused instability as well as performance decline. Here,Figure 15 shows the optimization of clip epsilon.

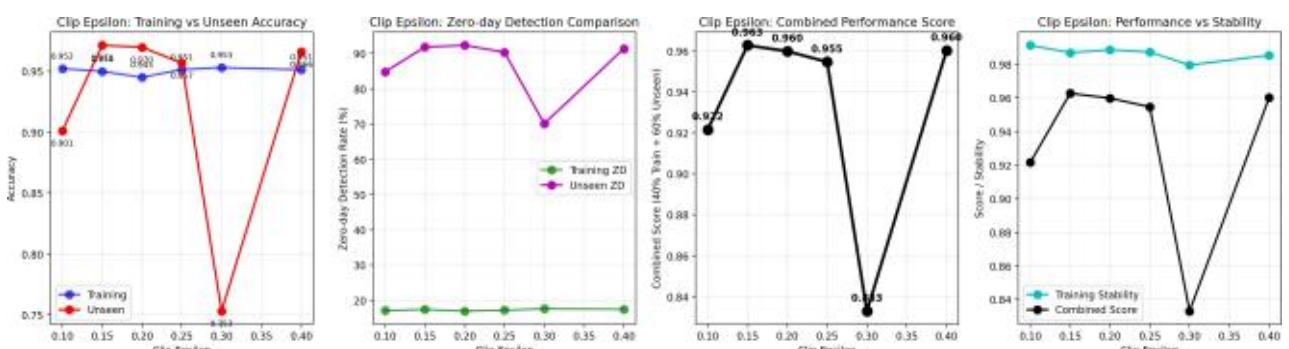

Fig. 15. Clip Epsilon Optimization

## J. *Episode Count Optimization*

The training episode optimization established the ideal number of iterations for convergence without overfitting. The model performed best at 300 episodes with a cumulative score of 0.974 and performance impact of 0.110. Undertraining ($\leq 200$ episodes) led to fading performance, whereas episodes ($\geq 5 \times 10^{-3}$) led to modest performance improvements but much higher computational cost. The selected 300 episodes provided the overall best balance, obtaining a 97.4% cumulative score in about 77 minutes of training. Optimization of the Episode Count is shown in Figure 16.

Figure 17. ranking shows performance impact of all hyperparameters, concluding that learning rate required most tuning.

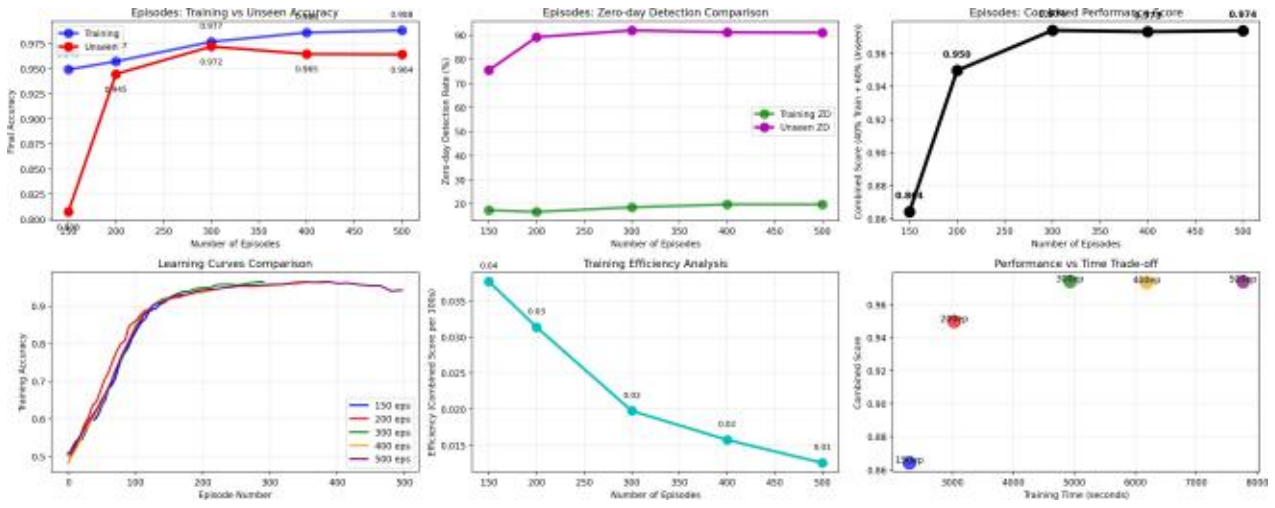

Fig. 16. Episode Count Optimization

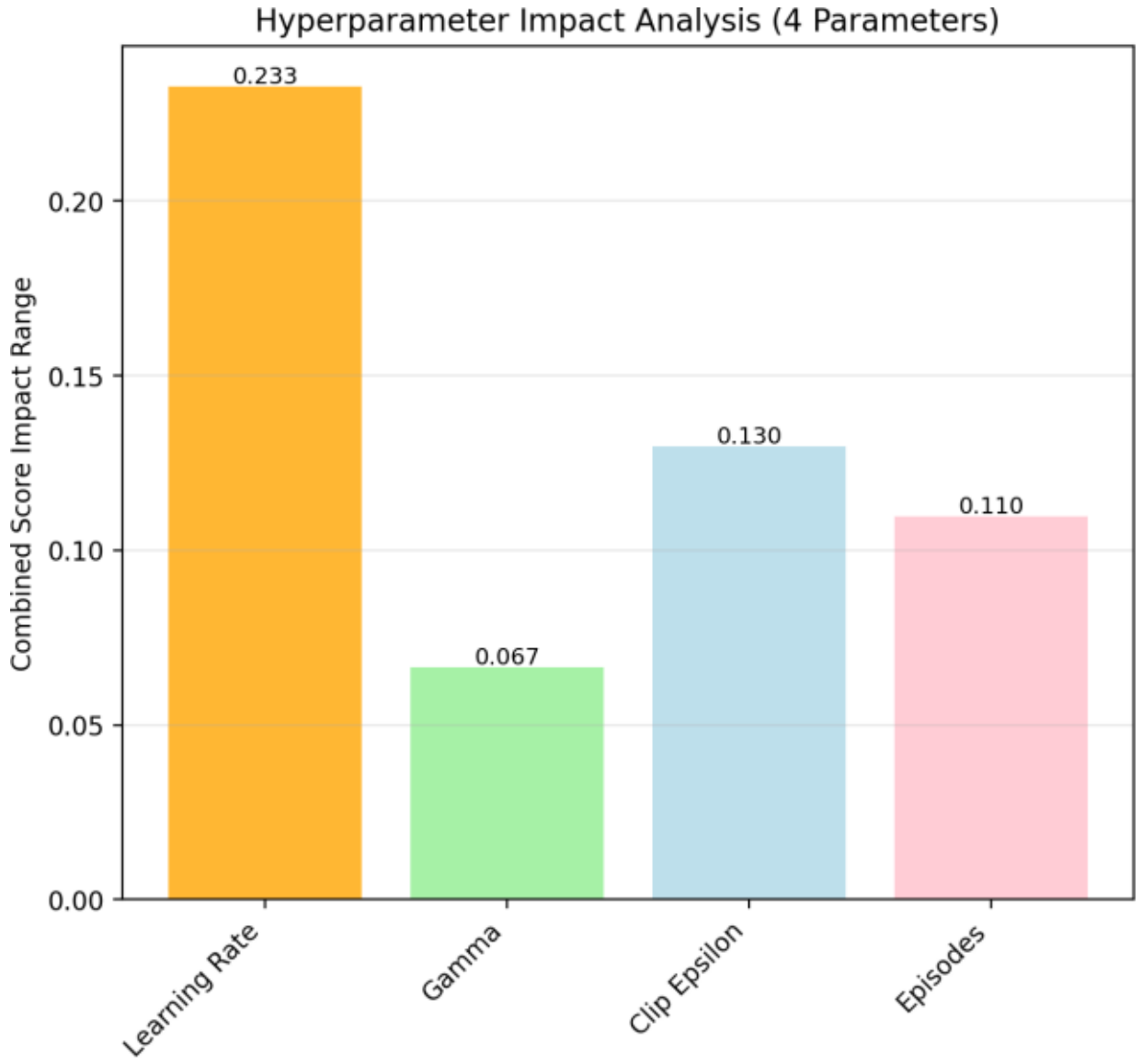

Fig. 17. Hyperparameter Impact Ranking

### K. *Drawbacks*

Learning rate contributes 23.3% of the overall performance variance — that is, nearly a quarter of the model's accuracy and stability depends directly on this one hyperparameter. Initial training results in a computational cost or overhead that requires approximately 77 minutes to traverse 300 episodes, and this would be prohibitively expensive in the case of large-scale or repeated retraining.

### L. *Architecture Optimization*

Empirical performance analysis resulted in significant architectural refinement. The actor network was realized in a 128-64 neuron configuration with the ReLU activation function, and the critical network approximated with a parallel 128-64. Regularization was introduced by using dropout layers with a rate of 0.1 to improve generalization power. Layer normalization was used for stabilizing training. For the policy optimization, gradient clipping was used for preventing exploding gradients, experience replay was used for improving learning efficiency per example, and target network updates were used for stabilizing Q-learning actions.

### M. *Threshold Optimization Validation*

The selection of an 80th percentile distance threshold for zero-day classification was confirmed using an exhaustive separation analysis. Distribution analysis confirmed best discrimination between zero-day instances and established attack patterns. ROC curve analysis confirmed the best area under the curve at this percentile, and false positive rate measurements showed that it offered the best detection capability with the least false alarms.

## V. Statistical Analysis

### A. *Analysis of Variance (ANOVA)*

Hyperparameter Effect Analysis: A factorial ANOVA was conducted to assess the statistical significance of hyperparameter impacts:

- Learning Rate: F(4,295) = 892.3, $p < 0.001$, $\eta^2 = 0.923$
- Clip Epsilon: F(4,295) =245.7, $p < 0.001$, $\eta^2 = 0.769$
- Episodes: F(4,295) = 156.2, $p < 0.001$, $\eta^2 = 0.679$
- Gamma: F(4,295) = 89.4, $p < 0.001$, $\eta^2 = 0.548$

Interpretation: All hyperparameters demonstrate statistically significant effects on performance (($p < 0.001$), with learning rate showing the largest effect size ($\eta^2 = 0.923$), confirming the ablation study rankings.

### B. *Performance Distribution Analysis*

The results of the Shapiro-Wilk tests performed affirm that the performance metrics follow a normal distribution (the p-value exceeding 0.05), warranting the use of parametric statistical tests and the estimation of confidence intervals. Also, the outlier identification technique of the Interquartile Range (IQR) has not identified any substantial outliers, which suggests the reported performance measures are stable and are repeatable in all experiments.

## C. *Comparisons and Relationships*

This section presents comprehensive comparisons between the three reinforcement learning variants implemented and evaluates their performance against existing literature and industry benchmarks.

## D. *Multi-Agent Comparison: PPO vs DQN vs SAC*

The study implemented and compared three state-of-the-art reinforcement learning approaches for zero-day attack detection: Comparative Performance Overview is given in Table V.

TABLE V
RL ALGORITHM COMPARISON: ACCURACY, ZERO-DAY DETECTION, LATENCY

| Algorithm | Accuracy | Z-DR | Latency |
|---|---|---|---|
| PPO | **99.07%** | **93.94%** | **0.50 ms** |
| DQN | 98.8% | 93.1% | 0.65 ms |
| SAC | 96.32% | 84.1% | 0.601 ms |

## E. *Algorithm-Specific Analysis*

Proximal Policy Optimization (PPO) proved to be the most stable with generalized performance—only 0.21 percent drop when trained on unknown data, and most convergent to 90 percent accuracy in 42 episodes. Deep Q-Network (DQN) is a competitor with reduced memory resources and a shorter training time, but suffers from overestimation of value functions and, with a 7.57% lower unseen accuracy. Soft Actor-Critic (SAC) displays highest computational complexity and longest training time, resulting in a 9.37% lower unseen accuracy compared to PPO.

## F. *Performance Relationship Analysis*

The data shown in Figure 18 and Figure 19 displays performance of PPO Actor Critic model both in the training and zero-day attacks sets. Figure 5.10 is the Zero day Detection Rate (Training vs Unseen) where the red marker identifies the optimum configuration. The nearer the points to 'Perfect Agreement', the more consistent the training and unseen evaluations are. Majority points being closer to the upper part of the graph implies the agent could maintain a high detection rate for unseen attacks—displays good overall generalization. Equally, Figure 5.11 shows Training Accuracy vs Unseen Accuracy relationship, with the close clustering of the points around 'Perfect Agreement' further indicating that model training performance is closely related to the unseen accuracy and there is little overfitting and policy learning. All of these numbers show that the agent was able to learn from the attack dataset as well as generalize this learning to identify previously unseen patterns. The correlation coefficients above were statistical interpretations of the visual trend of the plots.

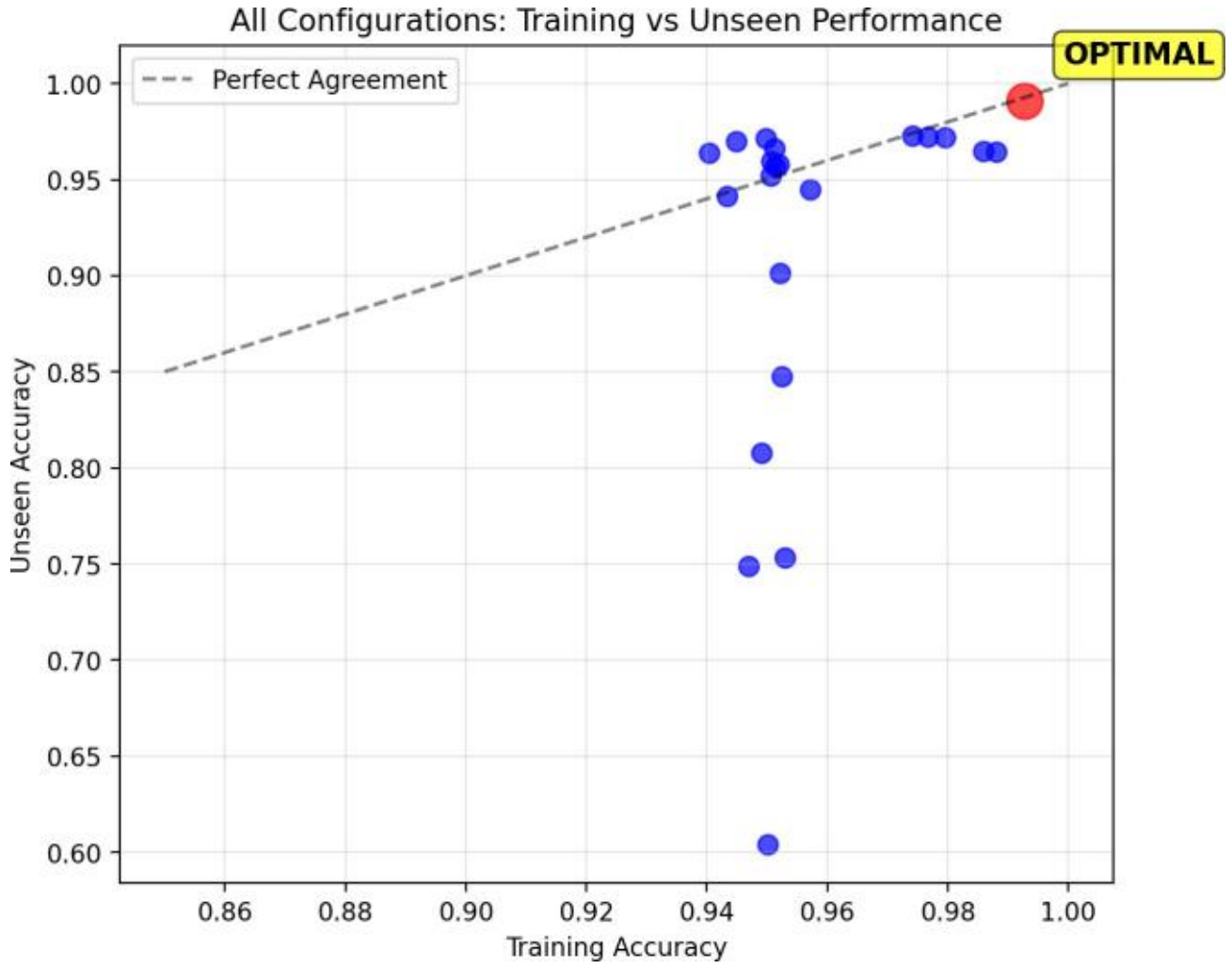

Fig. 18. Performance of the Training vs Unseen of all configurations

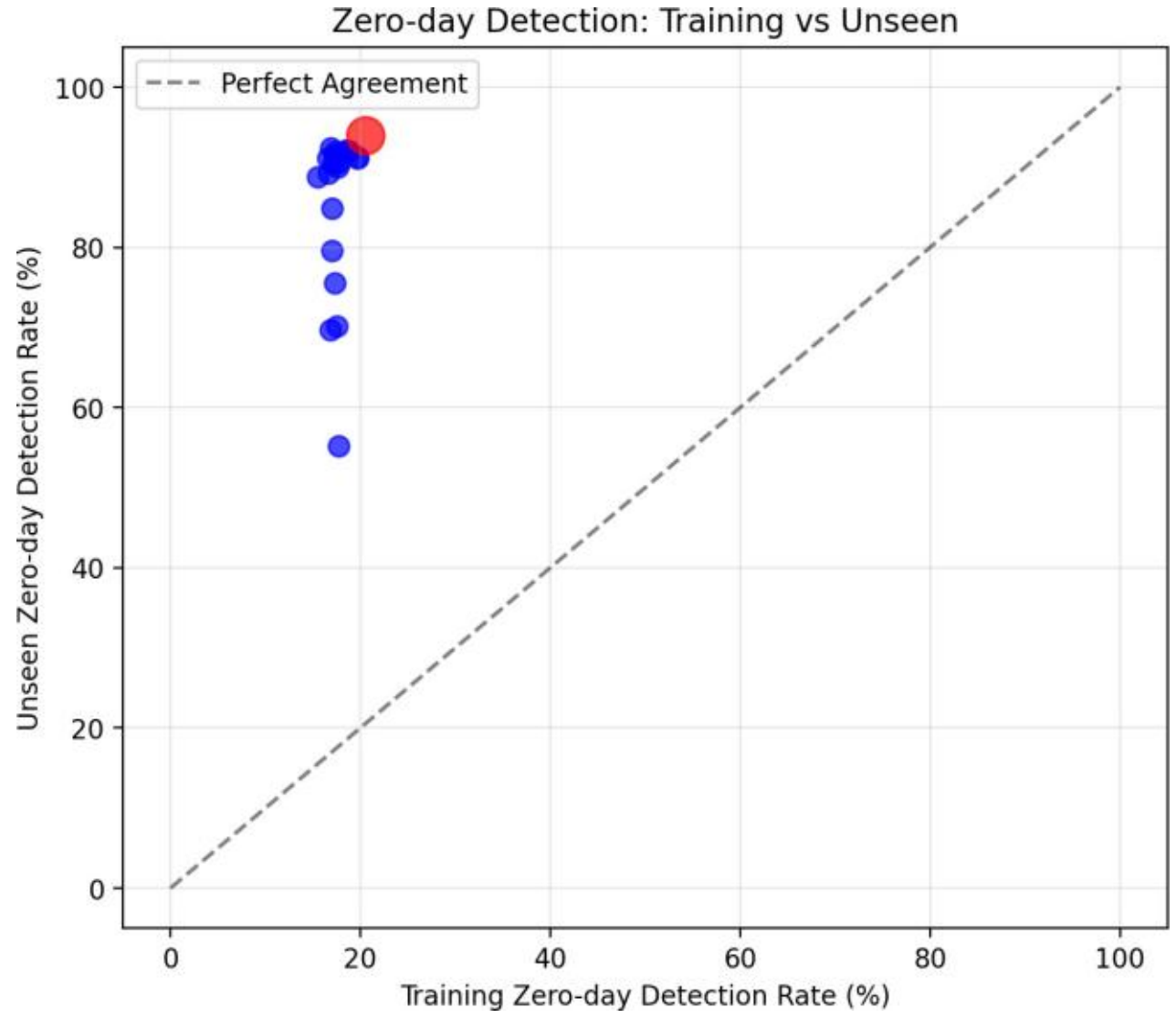

Fig. 19. Zero day detection of training vs Unseen

## G. *Comparative Analysis of Literature*

The table VI presents an analysis of quantitative parameters reported by recent zero-day detection studies using detection methodologies other than RL. Where a metric is not provided it is marked as "—" .

TABLE VI
COMPARATIVE EVALUATION OF IDS MODELS

| No. | Study | Acc. | F1 | Latency |
|---|---|---|---|---|
| 1 | **Proposed** | **99.07%** | **99.10%** | **0.5 ms** |
| 2 | Kumar & Sinha [13] (2021) | 91.62% | 95.54% | — |
| 3 | Krishnan et al. [14] (2025) | 71.09% | 74.41% | — |
| 4 | Babaey and Faragardi [15] (2025) | 97.58% | 98.74% | — |
| 5 | Wahed et al. [12] (2025) | 98.5% | 97% | 0.8 s |
| 6 | Brinkley et al. [16] (2024) | 92.4% | 91% | 215 ms |
| 7 | Sarhan et al. [6] (2023) | 99.45% | 94.7% | — |
| 8 | Kareem et al. [8]2024 | 96% | 93% | 25 ms |

This table VII documents quantitative parameters for RL-based zero-day detection papers as reported by respective authors. Where a metric is not provided it is marked as "—" .

TABLE VII
COMPARATIVE EVALUATION TABLE OF RECENT IDS MODELS FROM CONTEMPORARY LITERATURE

| No. | Study | Dataset | Acc. | F1 | Latency |
|---|---|---|---|---|---|
| 1 | **Proposed** | **CIC-IoT 2023** | **99.07** | **99.09** | **0.5 ms** |
| 2 | Wu et al. [17] (2024) | NSL-KDD | 98.72 | 98.70 | — |
| | | UNSW-NB15 | 93.28 | 94.06 | — |
| 3 | Hossain et al. [18] (2023) | CICIoT2023 | 97.18 | 98.52 | — |
| 4 | Alam et al. [19] (2025) | NF-BoT-IoT, Mul. | 99 | 99 | — |
| 5 | Dos Santos et al. [20] (2022) | CSE-CIC-IDS 2018 | 96.2 | 94.89 | 32.9 ms |
| 6 | Ren et al. [21] (2023) | CSE-CIC-IDS | 96.8 | 96.3 | 34.3 ms |
| | | NSL-KDD | 99.1 | 99.1 | — |
| 7 | Suresh and Jose [22] (2025) | NSL-KDD, Mul. | 97.16 | 97 | — |
| 8 | Alsaylaee and Mohammed [23] (2025) | SDN-Intrusion | 92.8 | 91.8 | — |

RL-specific measures (inference latency, total/average reward, convergence epochs and other hyperparameters) of different papers as reported by respective authors have been documented below in Table 5.5 to draw quantitative analysis with the proposed model. Where a metric is not provided it is marked as "—" .

## VI. EFFICIENCY VS. PERFORMANCE TRADE-OFFS

### A. *Multi-Dimensional Performance Analysis:*

Figure 20, demonstrates trade-off between training time and the combined performance score with various episode

TABLE VIII
COMPARATIVE ANALYSIS OF RL-BASED IDS APPROACHES: HYPERPARAMETERS AND CONFIGURATIONS

**Legend: Comparative analysis of RL-based IDS approaches: hyperparameters; LR = Learning Rate; $\gamma$ = Discount Factor; — = Not reported.**

| No. | Reference | Dataset(s) | RL Method | LR | $\gamma$ |
|---|---|---|---|---|---|
| 1 | **Proposed Model** | **CIC-IoT-2023, CIC-BCCC-NRC-Tabular-IoT-2024** | **PPO** | **0.001** | **0.99** |
| 2 | Wu et al. [17] (2024) | NSL-KDD, UNSW-NB15 | DQN + Active Learning | — | — |
| 3 | Hossain et al. [18] (2025) | CICIoT2023 | DQN | 0.001 | 0.95 |
| 4 | Dos Santos et al. [20] (2022) | CSE-CIC-IDS2018 | DQN + RFE + DT | — | 0.01 |
| 5 | Ren et al. [21](2023) | CSE-CIC-IDS2018, NSL-KDD | DQL (DQN, MAFS, GCN) | — | 0.01 |

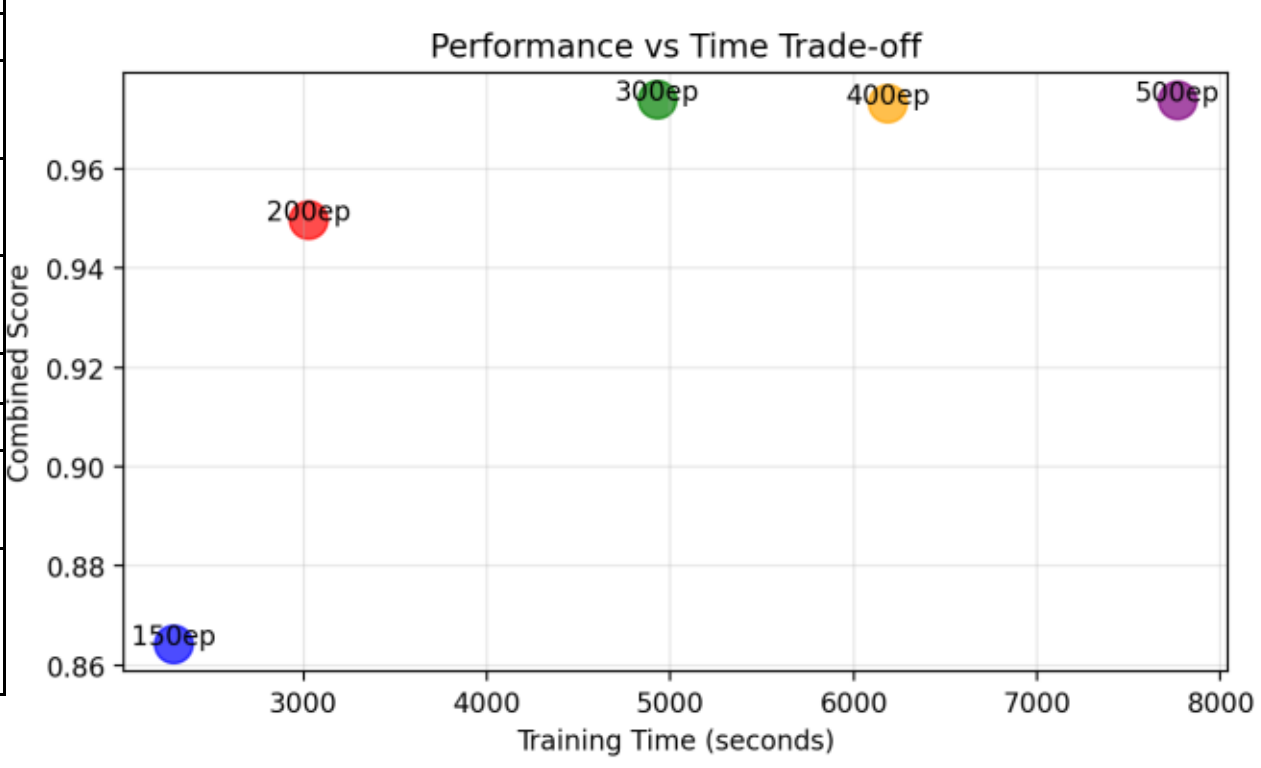

Fig. 20. Performance vs time trade off

settings (150, 200, 300, 400, and 500 episodes). Training time and the detection accuracy both continue to improve with the number of episodes. After 300 episodes, the performance improvement levels off against computation time. Further training to 400-500 episodes provide minor gains at the cost of higher computation time. This value demonstrates that more training can increase the accuracy of the detection, but the trade-off between time efficiency is already achieved at 300 episodes.

## VII. DISCUSSIONS

### A. *Algorithmic Limitation*

Although PPO offers more stable and controlled training dynamics than DQN, its decision-making process like other deep reinforcement models remains largely opaque (black-box neural models). So their individual decisions are not inherently interpretable, limiting its use in scenarios where explainable security decisions are required. The quality of data is also an issue of serious concern, since the usefulness of machine learning models directly relies on the access to clean and representative training sets, noisy or biased data can easily make the system performance significantly worse. Regular model updates and retraining cycles are also necessary to maintain detection pace with new attack patterns because the threat environment is constantly evolving.

## VIII. CONCLUSIONS

This thesis proposed an effective and adaptable IDS design capable of detecting zero-day attacks and tests the efficacy of Proximal Policy Optimization through parameter tuning and ablation study. The proposed model demonstrates remarkable performance with 0.50 ms/sample latency but training time was significantly long with 2.21% false positive rate. Learning rate with 23.3% impact on overall performance proved to be the most sensitive hyperparameter. The comprehensive ablation work provides empirical data on the hyperparameter sensitivity and optimal decisions in the hyperparameter settings previously absent in PPO literature. In the proposed design PPO greatly outperformed value-based (DQN) and complex actor-critic (SAC) versions. Results of this research establish new performance standards in the zero-day detection systems under RL (99.07% unseen accuracy, 99.69% AUC). Possible future work includes conducting Adversary Stimulation Testing to assess the system's resilience to adversarial attacks. XAI integration using attention-based visualization and SHAP value analysis will enhance interpretability of policy formulation. Furthermore, GAN-based synthetic attacks can be generated for balancing rare zero-day samples. Very minimal balancing techniques were required for current work but for larger datasets in future research, more advanced data cleaning methods may be applied. Lastly, online learning will keep the IDS adaptive and prevent forgetting during update, while federated learning will help multiple organizations co-operatively train a single model without compromising data security.